\documentclass{article}

\PassOptionsToPackage{numbers,compress}{natbib}

\usepackage[preprint]{neurips_2024}

\usepackage[utf8]{inputenc} 
\usepackage[T1]{fontenc}    
\usepackage{hyperref}       
\usepackage{url}            
\usepackage{booktabs}       
\usepackage{amsfonts}       
\usepackage{amsmath}
\usepackage{amssymb}
\usepackage{nicefrac}       
\usepackage{microtype}      
\usepackage{xcolor}         
\usepackage{graphicx}
\usepackage{tikz}
\usepackage{enumitem}

\newcommand{\vt}{\textsc{VibeTensor}}
\newcommand{\code}[1]{\texttt{#1}}

\title{VibeTensor: System Software for Deep Learning,\\\mbox{Fully Generated by AI Agents}}

\author{%
  \textbf{Bing Xu, Terry Chen, Fengzhe Zhou, Tianqi Chen, Yangqing Jia, Vinod Grover,}\\
  \textbf{Haicheng Wu, Wei Liu, Craig Wittenbrink, Wen-mei Hwu, Roger Bringmann,}\\
  \textbf{Ming-Yu Liu, Luis Ceze, Michael Lightstone, Humphrey Shi}\\
  NVIDIA\\
}

\begin{document}

\maketitle

\begin{abstract}
\vt{} is an open-source research system software stack for deep learning, generated by LLM-powered coding agents under high-level human guidance. In this paper, ``fully generated'' refers to code provenance: implementation changes were produced and applied as agent-proposed diffs; validation relied on builds, tests, and differential checks executed by the agent workflow, without per-change manual diff review. It implements a PyTorch-style eager tensor library with a C++20 core (CPU+CUDA), a torch-like Python overlay via nanobind~\cite{nanobind}, and an experimental Node.js/TypeScript interface. Unlike thin bindings, \vt{} includes its own tensor/storage system, schema-lite dispatcher, reverse-mode autograd engine, CUDA runtime (streams/events/graphs~\cite{cuda_graphs}), a stream-ordered caching allocator with diagnostics, and a stable C ABI for dynamically loaded operator plugins. We view this open-source release as a milestone for AI-assisted software engineering: it demonstrates that coding agents can generate a coherent deep learning runtime spanning language bindings down to CUDA memory management, with validation constrained by builds and tests. We describe the architecture, summarize the workflow used to produce and validate the system, and evaluate the resulting artifact. We report repository scale and test-suite composition, and summarize reproducible microbenchmarks from an accompanying AI-generated kernel suite, including fused attention measured against PyTorch SDPA/FlashAttention~\cite{flashattention}. We also report end-to-end training sanity checks on three small workloads (sequence reversal, CIFAR-10 ViT, and a miniGPT-style model) on NVIDIA H100 (Hopper, SM90) and Blackwell-class GPUs; multi-GPU results are Blackwell-only and rely on an optional CUTLASS-based ring-allreduce plugin gated on CUDA 13+ and \code{sm103a} toolchain support. Finally, we discuss failure modes that arise in generated system software---in particular a ``Frankenstein'' composition effect where locally correct subsystems can interact to produce globally suboptimal performance. We open-source the resulting system software and evaluation artifacts at \url{https://github.com/NVLabs/vibetensor}.
\end{abstract}

\section{Introduction}
Deep learning systems such as TensorFlow~\cite{tensorflow_paper} and PyTorch~\cite{pytorch_paper} have matured into large software stacks spanning language frontends, runtime systems, device libraries, compilers, and kernels.
Building and evolving such systems has historically required sizable teams and long time horizons.
At the same time, recent progress in large language models (LLMs) trained on code~\cite{codex} and tool-using agents~\cite{react} has enabled workflows where models propose code changes and validate them using external tools (search, builds, and tests).

We ask a concrete systems question: \emph{can coding agents generate a coherent deep learning system software stack that crosses the usual abstraction boundaries---from Python and JavaScript APIs down to C++ runtime components and CUDA memory management---and validate it through builds and tests?}
We open-source \vt{} under the Apache 2.0 license as a technical artifact for answering this question empirically.
We do not introduce a new agent framework here; instead, we treat agents as black boxes and focus on the released system software and the validation scaffolding that constrains system-scale generation.

\paragraph{Platforms.}
The release targets Linux x86\_64 and NVIDIA GPUs via CUDA; builds without CUDA are intentionally disabled.
In this paper we report runs on NVIDIA H100 (Hopper, SM90) and Blackwell-class GPUs.
Multi-GPU results are Blackwell-only and rely on the experimental Fabric subsystem together with an optional CUTLASS-based ring-allreduce plugin; the plugin is best-effort and gated on CUDA 13+ / \code{sm103a} toolchain support.

\paragraph{Scope.}
The goal of this paper is \emph{not} to introduce a new training framework or to claim state-of-the-art end-to-end performance.
Instead, we focus on (1) the architecture of the generated system, (2) the methodology used to produce and validate it, and (3) lessons for future ``vibe-coded'' system software.

\paragraph{Contributions.}
\begin{itemize}[leftmargin=*]
  \item \textbf{End-to-end system software.} We describe a deep learning runtime with multi-language frontends (Python and Node.js) backed by a shared C++ dispatcher, tensor/storage implementation, autograd engine, and CUDA subsystem.
  \item \textbf{Interoperability and extension points.} We document DLPack interop~\cite{dlpack}, a stable C plugin ABI, and hooks for custom kernels authored in Triton~\cite{triton} and CUDA template libraries such as CUTLASS~\cite{cutlass}.
  \item \textbf{AI-assisted development methodology.} We summarize a practical workflow for generating and validating system-scale software with agents, using builds, tests, differential checks, and multi-agent code review as guardrails.
  \item \textbf{Artifact evaluation.} We report repository scale and test composition, summarize reproducible microbenchmarks for an accompanying AI-generated kernel suite (including attention compared against PyTorch SDPA/FlashAttention), and report end-to-end training sanity checks (sequence reversal, CIFAR-10 ViT, and miniGPT) plus a Blackwell-only multi-GPU scaling benchmark.
  \item \textbf{Lessons and limitations.} We describe failure modes observed in generated system software, including a ``Frankenstein'' composition effect where correctness-first subsystems can structurally inhibit performance.
\end{itemize}

\section{Background and related work}
\label{sec:related}
\paragraph{Eager execution and define-by-run.}
Define-by-run systems such as Chainer~\cite{Chainer} and DyNet~\cite{DyNet} established an imperative approach to building dynamic computation graphs.
PyTorch~\cite{pytorch_paper} popularized this model with a Python-first interface and a performance-oriented C++ core.
\vt{} follows the same eager execution philosophy, but our emphasis is on the feasibility of generating a multi-layer stack (frontends, runtime, CUDA allocator/graphs) with coding agents.

\paragraph{Kernel authoring systems.}
Recent systems such as Triton~\cite{triton} make custom kernel authoring accessible in a Python-native workflow.
CUDA template libraries such as CUTLASS~\cite{cutlass} provide a complementary approach for hand-optimized custom kernels.
Throughout the paper, CUTLASS refers to the C++ template library; CuTeDSL refers to a separate DSL backend used in our accompanying kernel suite.
\vt{} includes integration hooks for both styles and ships a kernel suite that exercises these pathways.

\paragraph{AI-assisted software engineering.}
Code-focused LLMs~\cite{codex} and tool-using agents~\cite{react} have enabled workflows where software artifacts are constructed by iteratively proposing edits and validating them via tools.
Benchmarks such as SWE-bench~\cite{swebench} evaluate issue-resolution capability in real repositories.
\vt{} differs in that the artifact spans system-software layers down to CUDA memory management, and the open-source release is intended as a substrate for studying how AI-assisted workflows behave at system scale.

\section{Design principles}
\label{sec:design_principles}
\vt{} targets a pragmatic subset of the design space.
The principles below shaped both the architecture and the evaluation criteria used during generation.

\paragraph{Eager execution with explicit control.}
\vt{} follows the define-by-run model popularized by PyTorch~\cite{pytorch_paper}: operator calls execute immediately, and autograd traces a dynamic graph.
This model is well matched to debugging, rapid iteration, and dynamic control flow.

\paragraph{End-to-end coherence.}
The system was generated to span the full path from user-facing APIs to GPU execution.
We prioritize having \emph{working} implementations of tensors, dispatch, autograd, and CUDA runtime mechanisms, even when incomplete or unoptimized.

\paragraph{Testability as a first-class constraint.}
In a generated codebase, tests serve as executable specifications and regression guards.
\vt{} includes C++ (CTest) and Python (pytest) test suites, as well as an import-only API-parity checker against PyTorch for a scoped manifest.
The Python CUDA surface includes contract-tested runtime and allocator APIs (e.g., \code{memory\_stats}, \code{memory\_snapshot}, and per-process memory-fraction caps).
We do not claim fully automated test synthesis; rather, the tests are included in the released artifact and can be inspected and rerun.
In this project, tests were written in the same agent loop as the implementation: when failures were discovered, the workflow typically responded by adding a minimal regression test or contract check and rerunning the suite.
This is still hand-written test code (not formal synthesis), and test authoring/maintenance remains a scalability cost as the API surface grows.

\paragraph{Interoperability and extensibility.}
We treat the runtime as a component in a larger ecosystem.
\vt{} supports zero-copy exchange via DLPack~\cite{dlpack}, provides a C ABI for plugins, and exposes Python-level override mechanisms for rapid experimentation.

\paragraph{Inspectability.}
Generated systems benefit from being observable.
\vt{} therefore includes allocator diagnostics implemented directly in its CUDA caching allocator---per-device statistics and segment/pool snapshot APIs exposed through the Python surface (e.g., \code{memory\_stats} and \code{memory\_snapshot})---along with CUDA graph instrumentation and optional multi-GPU observability hooks (Section~\ref{sec:fabric}).
These diagnostics are native to \vt{}: the API is PyTorch-inspired, but the counters and snapshots reflect \vt{}'s allocator state rather than calling into PyTorch.

\section{System overview}
\label{sec:overview}
\begin{figure}[t]
  \centering
  \includegraphics[width=\linewidth]{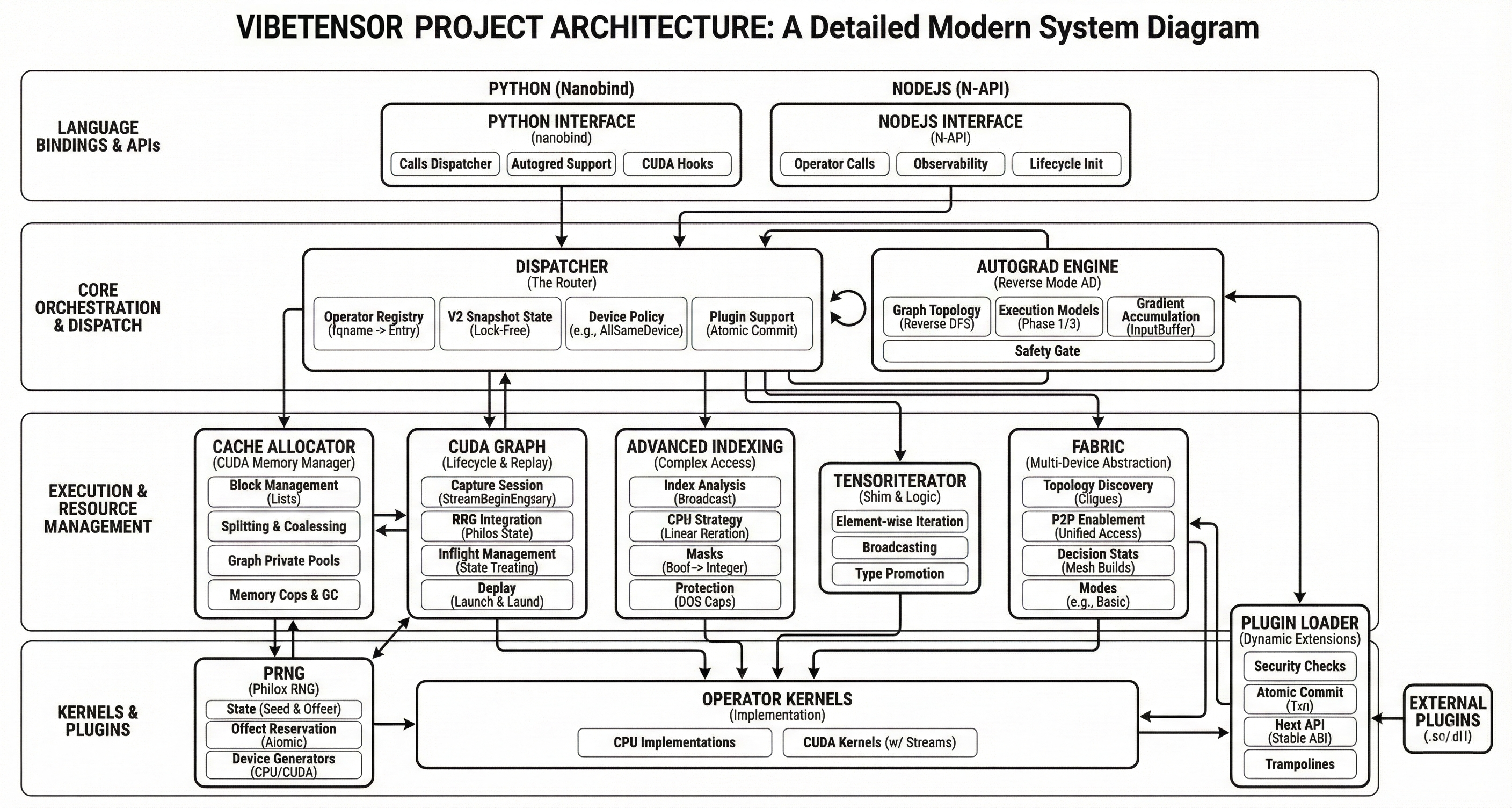}
  \caption{High-level architecture of \vt{}: Python (nanobind) and Node.js (N-API) frontends dispatch into a shared C++ core implementing tensors/storage, dispatch, autograd, indexing, RNG, and CUDA runtime components (streams/events/graphs and caching allocator). Optional kernel libraries and dynamically loaded plugins extend the operator set.}
  \label{fig:arch}
\end{figure}

Figure~\ref{fig:arch} shows the macro architecture.
\vt{} is organized into frontends, a shared core runtime, a CUDA subsystem, and optional kernel and plugin layers.

\subsection{Frontends: Python and Node.js}
\label{sec:frontends}
\vt{} exposes a torch-like Python overlay (\code{vibetensor.torch}) implemented on top of a nanobind~\cite{nanobind} extension module.
The overlay provides tensor factories, an \code{ops} namespace that routes to the dispatcher, and CUDA utilities for streams, events, allocator stats, and CUDA graph capture.

\vt{} also includes an experimental Node.js addon based on Node-API~\cite{node_api}.
The JavaScript/TypeScript (JS/TS) overlay is async-first: heavy work is scheduled via \code{napi\_async\_work} to avoid blocking the Node event loop, and a process-wide inflight cap (\code{VBT\_NODE\_MAX\_INFLIGHT\_OPS}) bounds queued work.
In the current prototype, JavaScript tensor objects are CPU-only; CUDA is exposed through explicit H2D/D2H and DLPack import/export.

\subsection{Core runtime: tensors, storage, and views}
\label{sec:tensors}
The C++ core defines a \code{TensorImpl} as a view over reference-counted \code{Storage}, with sizes/strides, storage offset, dtype, and device metadata.
This design supports non-contiguous views and as\_strided semantics, and maintains an atomic version counter (shared by views) to support in-place safety checks.

For elementwise and reduction operators, \vt{} provides a TensorIterator subsystem that computes iteration shapes and per-operand stride metadata.
TensorIterator is also exposed through the plugin ABI so external kernels can reuse iteration logic safely.

\subsection{Dispatcher: schema-lite operator registry}
\label{sec:dispatcher}
\vt{} implements a schema-lite dispatcher that maps fully qualified operator names (e.g., \code{vt::add}) to kernel implementations.
The dispatcher supports CPU and CUDA dispatch keys, boxed and unboxed call paths, and wrapper layers (e.g., autograd and Python overrides).
Registration publishes immutable per-operator snapshot states that encode base kernels and wrapper presence, enabling lock-free call paths in steady state.
Device-policy rules enforce common invariants (such as ``all tensor inputs on the same device'') while allowing specialized policies for experimental multi-device paths.

\subsection{Reverse-mode autograd}
\label{sec:autograd}
Autograd is implemented as a reverse-mode engine with \code{Node}/\code{Edge} graph objects and per-tensor \code{AutogradMeta}.
During backward, the engine maintains dependency counts, input buffers, and a ready queue.
For CUDA tensors, the engine records and waits on CUDA events to synchronize cross-stream gradient flows.
\vt{} also contains an experimental multi-device mode intended for research on cross-device graph execution.

\subsection{CUDA subsystem: streams, allocator, and graphs}
\label{sec:cuda}
\vt{} provides C++ wrappers for CUDA streams and events, a caching allocator with stream-ordered semantics, and CUDA graph capture/replay support~\cite{cuda_graphs}.
The allocator includes diagnostics (snapshots, statistics, fraction caps, and GC ladders) to make memory behavior observable during testing and debugging.
CUDA graphs integrate with allocator ``graph pools'' to manage memory lifetime across capture and replay.

\subsection{Interop and extensions}
\label{sec:extensions}
\vt{} supports interoperability via DLPack import/export~\cite{dlpack} for both CPU and CUDA tensors.
It also includes a C++20 safetensors loader/saver interface for efficient serialization~\cite{safetensors}.

For extensibility, \vt{} offers:
(i) Python-level overrides inspired by \code{torch.library},
(ii) a stable C ABI for dynamically loaded plugins that register new operators and kernels, and
(iii) hooks for integrating custom GPU kernels authored in domain-specific systems such as Triton~\cite{triton} or CUDA template libraries such as CUTLASS~\cite{cutlass}.
The C plugin ABI is versioned and exposes DLPack-based dtype/device metadata and TensorIterator helpers, allowing external kernels to reuse iteration logic and obey aliasing rules.

\subsection{Fabric and experimental multi-GPU components}
\label{sec:fabric}
\vt{} includes an experimental \emph{Fabric} subsystem that exposes explicit peer-to-peer GPU access paths (CUDA P2P and UVA when supported by the device topology) and observability (stats and event snapshots).
The intent is to enable research on single-process multi-GPU execution where the runtime can reason about topology, synchronization, and memory placement.
In the current release, Fabric focuses on explicit elementwise operations and observability primitives rather than a full distributed training runtime.

\begin{figure*}[t]
  \centering
  \includegraphics[width=0.98\textwidth]{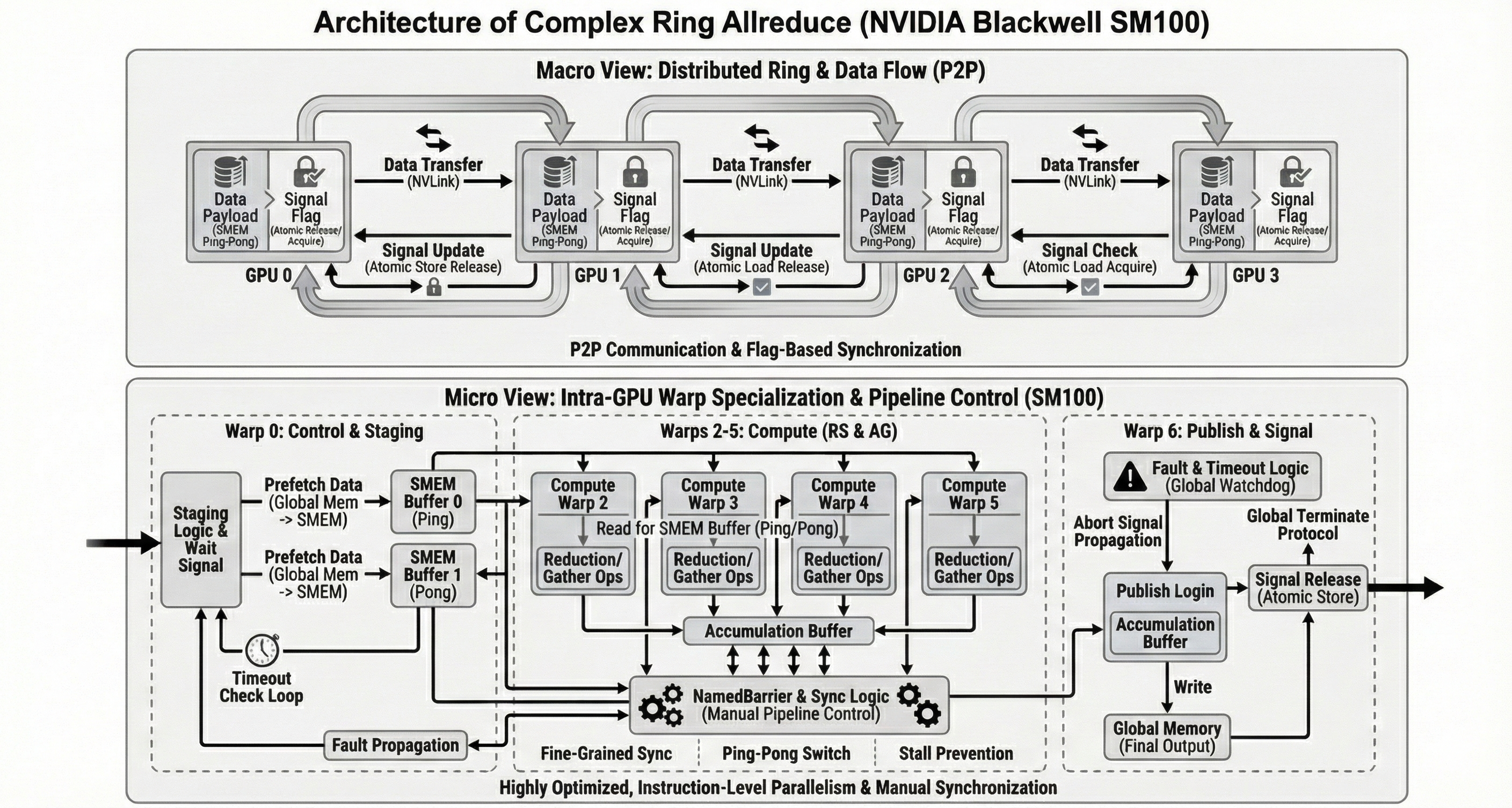}
  \caption{Macro and micro view of a warp-specialized ring allreduce kernel (Blackwell SM100/SM103) shipped as an example plugin. The plugin is best-effort, gated on toolchain and hardware availability, and is not required for using \vt{}'s core runtime; it is intended as a reference implementation rather than a replacement for NCCL.}
  \label{fig:ring_allreduce}
\end{figure*}

As a reference example of hardware-specialized extensions, the release also includes a best-effort CUTLASS-based ring allreduce plugin targeting NVIDIA Blackwell-class GPUs; this is an illustrative plugin (not an NCCL-based drop-in replacement). The plugin directly binds CUTLASS experimental ring-allreduce kernels and does not call NCCL; for comparison, the plugin directory includes a standalone PyTorch \code{torch.distributed} (NCCL) bandwidth benchmark.
Figure~\ref{fig:ring_allreduce} illustrates the macro (inter-GPU ring) and micro (warp-specialized pipeline) view.

\section{AI-assisted development methodology}
\label{sec:method}
\vt{} was developed with LLM-powered coding agents under high-level human guidance over approximately two months.
Because the focus of this paper is system software rather than the agent implementation, we treat the agents as black boxes that propose code changes and use tools to validate them.
Humans provided high-level requirements and priorities, but did not perform diff-level review or run validation commands; instead, agents executed builds, tests, and differential checks and retained changes when these checks passed.

\subsection{Workflow and guardrails}
The development workflow repeatedly applied a simple loop:
(1) specify a scoped goal and invariants,
(2) generate and apply code changes,
(3) compile and run focused tests,
and (4) broaden validation as subsystems composed.

Two guardrails were especially important:
\textbf{tests as specifications} (CTest/pytest, plus targeted multi-step regression tests), and
\textbf{differential checks} against reference implementations (e.g., PyTorch for selected operators, DLPack roundtrips, and kernel-vs-kernel comparisons).
We also used multi-agent code review to catch missing cases, redundant abstractions, and unsafe patterns that a single agent might overlook.

\subsection{Case study: debugging across layers}
A recurring pattern in agent-generated system software is that ``single-shot'' correctness (an operator matches a reference once) does not imply stability when the subsystem is repeatedly composed inside a training loop.
Development logs from one attention-kernel porting episode illustrate the range of issues that arose: launch-time limits (kernel parameter counts, shared-memory usage), numerical subtleties (log base and scaling in stabilized attention), and runtime hazards such as uninitialized GPU buffers whose reuse only manifests after several iterations.

Table~\ref{tab:debug_cases} summarizes representative issues and the guardrails that prevented recurrence.

\begin{table}[t]
  \caption{Representative debugging episodes observed during agent-assisted development (selected from development notes).}
  \label{tab:debug_cases}
  \centering
  \small
  \begin{tabular}{p{0.22\linewidth} p{0.36\linewidth} p{0.32\linewidth}}
    \toprule
    Symptom & Root cause & Fix / regression guard \\
    \midrule
    Kernel crash or segfault & Exceeded kernel parameter limits (e.g., too many explicit stride arguments) & Simplify call interface; add launch-configuration tests \\
    Shared-memory compile failure & Requested $>48$KB of statically allocated shared memory (toolchain cap) & Adaptive tiling / autotune; compile-time guards \\
    Large numerical error vs reference & Incorrect stabilization math (e.g., log2 vs ln scaling in attention) & Use flash-attention-style formulation; differential tests vs PyTorch \\
    Training loss divergence after many steps & Uninitialized GPU buffers reused during gradient accumulation & Initialize gradient buffers; add multi-step training regressions \\
    \bottomrule
  \end{tabular}
\end{table}

\section{Evaluation}
\label{sec:evaluation}
We evaluate \vt{} along four axes: repository scale, correctness infrastructure, kernel-level performance of selected components, and end-to-end training sanity checks.

\subsection{Repository scale}
Table~\ref{tab:scale} summarizes the scale of the codebase in this release.
Counts exclude vendor-provided third-party dependencies.

\begin{table}[t]
  \caption{\vt{} repository scale (excluding vendor-provided third-party code). ``Core'' refers to \code{include/} and \code{src/} in the C++ runtime.}
  \label{tab:scale}
  \centering
  \begin{tabular}{lrr}
    \toprule
    Component & Source files & Non-blank LOC \\
    \midrule
    Core C++/CUDA runtime & 218 & 63,543 \\
    Plugins (C/CUDA) & 50 & 17,500 \\
    Python overlay & 33 & 9,016 \\
    Node.js/TS overlay & 17 & 2,010 \\
    Tests (C++) & 194 & 32,042 \\
    Tests (Python) & 225 & 21,913 \\
    AI kernel suite (Python, Triton/CuTeDSL) & 203 & 55,882 \\
    \bottomrule
  \end{tabular}
\end{table}

\subsection{Correctness and reproducibility}
\vt{} includes C++ unit tests (run via CTest) and Python tests (run via pytest).
The reference build targets Linux x86\_64 and requires a CUDA toolchain; builds without CUDA are intentionally disabled in the current release.
Tests cover core tensor semantics, dispatcher behavior, autograd, CUDA runtime utilities, allocator invariants, DLPack interop, CUDA graphs, and selected kernel paths.
Allocator diagnostics are surfaced via the Python CUDA module (\code{vibetensor.torch.cuda}), including \code{memory\_stats}, \code{memory\_snapshot}, and per-process memory-fraction caps; the Python test suite includes contract checks for these APIs.

To reduce ``it imports but it is broken'' failures, \vt{} also provides an import-only API-parity gate: a script that checks a scoped manifest of importable symbols against PyTorch.
The manifest is intentionally limited and should be viewed as a regression guard rather than a claim of full API compatibility.

\paragraph{Reproducing checks.}
A typical validation sequence is:
\begin{itemize}[leftmargin=*]
  \item \code{python -m pip install -U pip build pytest numpy}
  \item \code{CMAKE\_BUILD\_TYPE=Debug python -m pip install -v -e .[test]}
  \item \code{ctest --test-dir build-py -j\$(nproc) --output-on-failure}
  \item \code{pytest -q}
  \item \code{python tools/check\_api\_parity.py --manifest api/manifest\_import\_only.json}
  \item (optional) Node overlay: \code{cd js/vibetensor \&\& npm ci \&\& npm test}
\end{itemize}

\subsection{Kernel suite and microbenchmarks}
\begin{figure}[t]
  \centering
  \includegraphics[width=\linewidth]{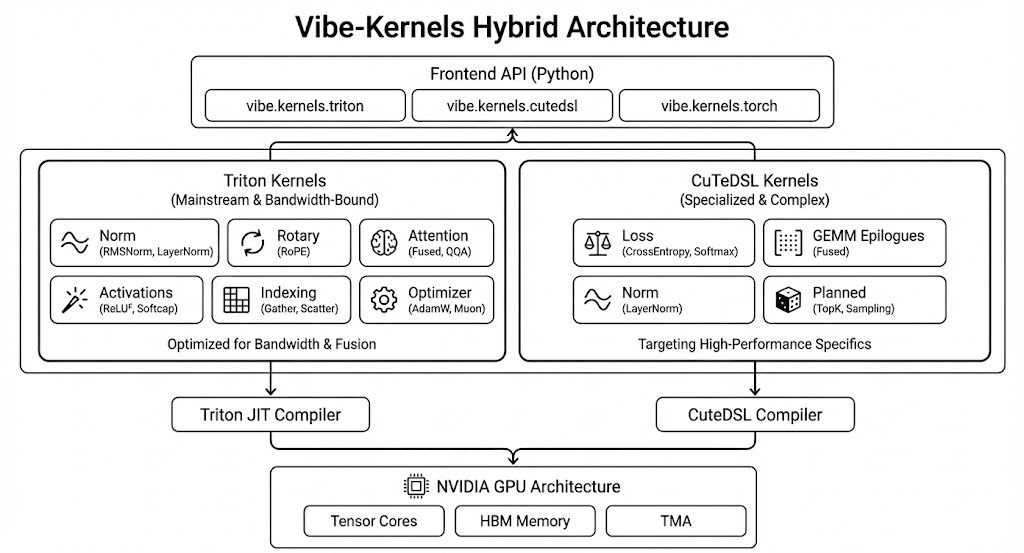}
  \caption{Macro view of the accompanying AI-generated kernel suite. Multiple backends (Triton, CuTeDSL, and PyTorch reference paths) share a common Python-facing interface and benchmark harnesses.}
  \label{fig:vibekernels}
\end{figure}

The accompanying AI-generated kernel suite provides Triton and CuTeDSL implementations for selected operators (e.g., norms, rotary embeddings, optimizers, and softmax/loss), along with benchmark harnesses that record performance and numerical diffs against PyTorch baselines.
Most Triton and CuTeDSL kernels in this suite are exercised via a PyTorch-facing harness (PyTorch tensors and CUDA device discovery), which also serves as the reference for differential checks.
Where available, we additionally provide \vt{}-native wrappers (typically via DLPack) that can run without importing PyTorch; this pathway currently covers only selected kernels.
We additionally include a dedicated fused-attention benchmark that compares a Triton kernel against PyTorch scaled dot-product attention (SDPA), which dispatches to FlashAttention when available.
Figure~\ref{fig:vibekernels} shows the macro structure of the kernel suite.

Table~\ref{tab:kernels} reports selected microbenchmarks extracted from kernel-suite reports and the fused-attention benchmark.
Because \vt{} is async-first, all timings synchronize around the measured region to reflect device time rather than host dispatch latency.

\paragraph{Benchmark environment.}
Unless otherwise specified, the reported timings were collected on NVIDIA H100 PCIe (SM90) systems under CUDA 12--13 with PyTorch 2.9 nightly (to match the SDPA/FlashAttention baseline in this environment) and Triton 3.5.0.
PyTorch is used here as a baseline and as a harness dependency for most Triton/CuTeDSL kernels; the \vt{} core runtime does not depend on PyTorch.
Each benchmark seeds RNGs, performs equal warmup for baseline and candidate, synchronizes around timed sections, and reports mean latency and numerical diffs (max absolute error and an allclose check).

\begin{table}[t]
  \caption{Selected microbenchmarks from the accompanying kernel suite and the fused-attention benchmark on H100 PCIe (recorded in release artifacts). \code{Torch} refers to the PyTorch baselines used by the harnesses (e.g., \code{torch.nn.functional.layer\_norm}, \code{torch.nn.functional.rms\_norm}, and \code{torch.nn.functional.scaled\_dot\_product\_attention} for attention/SDPA). \code{Best} reports the lowest measured latency among non-Torch backends (e.g., Triton, CuTeDSL) for that row. These results characterize isolated kernels, not end-to-end model performance.}
  \label{tab:kernels}
  \centering
  \small
  \begin{tabular}{llrrr}
    \toprule
    Kernel & Shape/Dtype & Torch (ms) & Best (ms) & Speedup \\
    \midrule
    LayerNorm (fwd+bwd) & (4096, 8192) (BF16) & 0.47 & 0.45 & 1.06$\times$ \\
    RMSNorm (fwd) & (4096, 8192) (BF16) & 0.82 & 0.13 & 6.3$\times$ \\
    Rotary (fwd) & (4, 8, 2048, 128) (BF16) & 0.73 & 0.14 & 5.33$\times$ \\
    Attention (fwd) & (32, 10, 10, 2048, 128) (BF16, causal) & 2.24 & 1.46 & 1.54$\times$ \\
    Attention (bwd) & (32, 10, 10, 2048, 128) (BF16, causal) & 8.78 & 6.97 & 1.26$\times$ \\
    \bottomrule
  \end{tabular}
\end{table}

\paragraph{Attention discussion.}
The attention benchmarks show a mixed picture.
For a NanoChat-style training configuration (batch 32, sequence length 2048), the Triton kernel outperforms the PyTorch SDPA/FlashAttention baseline (1.54$\times$ forward, 1.26$\times$ backward).
However, for small-batch GQA prefill workloads, the same pointer-based kernel can trail FlashAttention (0.67$\times$ forward, 0.66$\times$ backward).
We view this as a representative example of how performance portability depends on kernel specialization and toolchain maturity (e.g., availability of Hopper warp-specialization knobs in Triton).

\subsection{End-to-end training runs}
\begin{figure*}[t]
  \centering
  \begin{minipage}{0.32\textwidth}
    \centering
    \includegraphics[width=\linewidth]{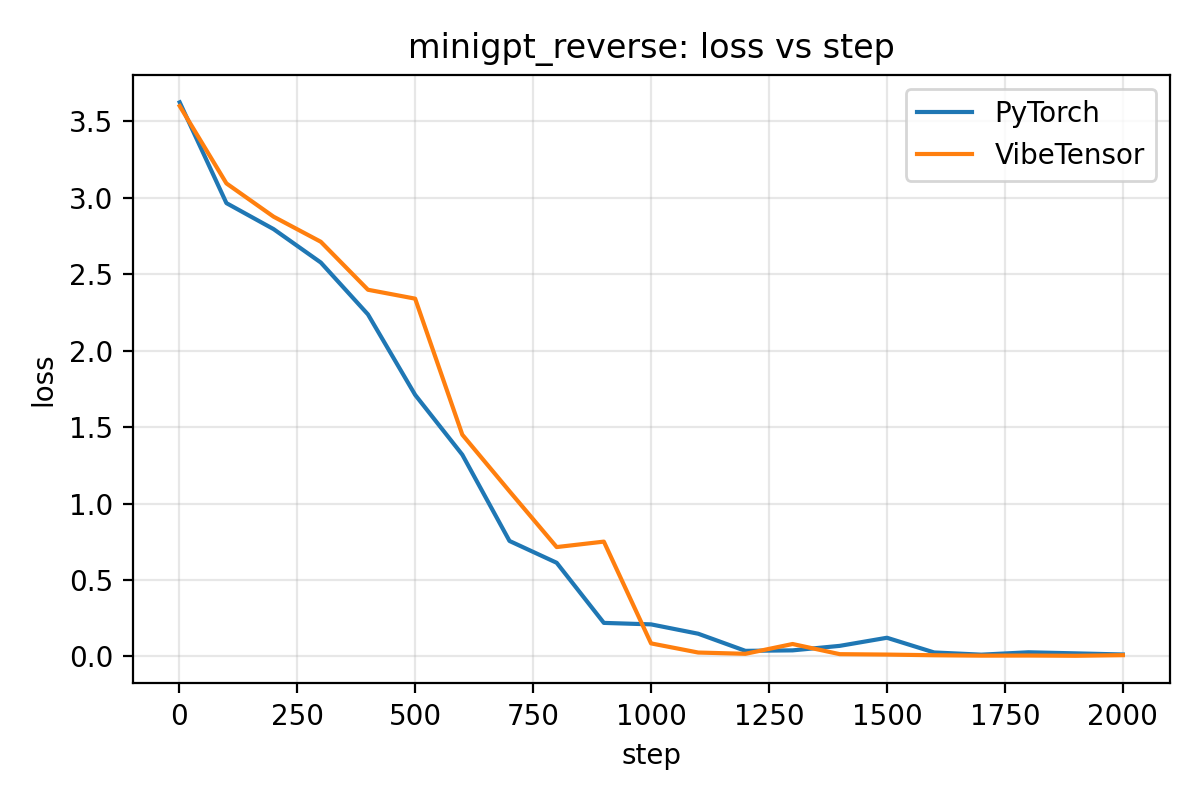}\\
    \small (a) Reversal: loss
  \end{minipage}
  \hfill
  \begin{minipage}{0.32\textwidth}
    \centering
    \includegraphics[width=\linewidth]{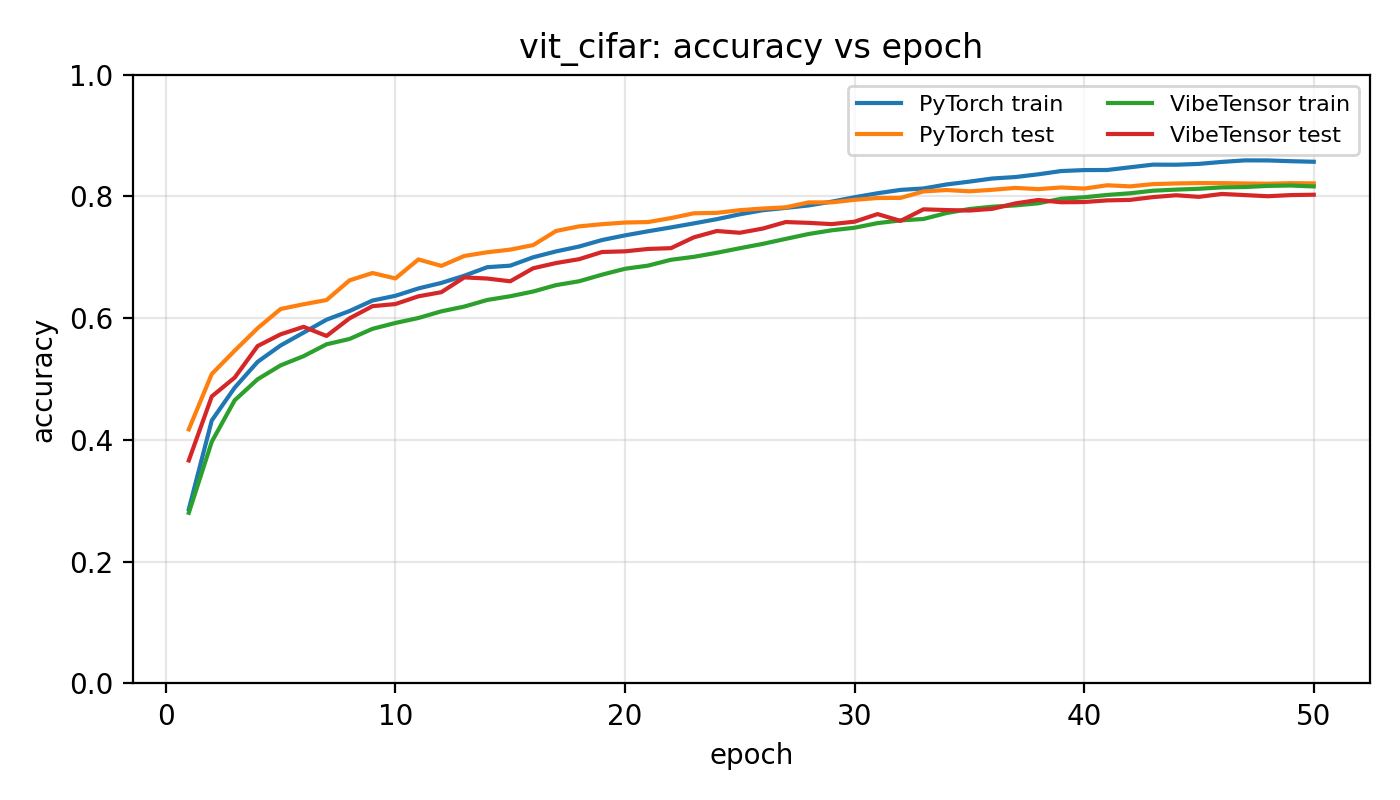}\\
    \small (b) ViT: accuracy
  \end{minipage}
  \hfill
  \begin{minipage}{0.32\textwidth}
    \centering
    \includegraphics[width=\linewidth]{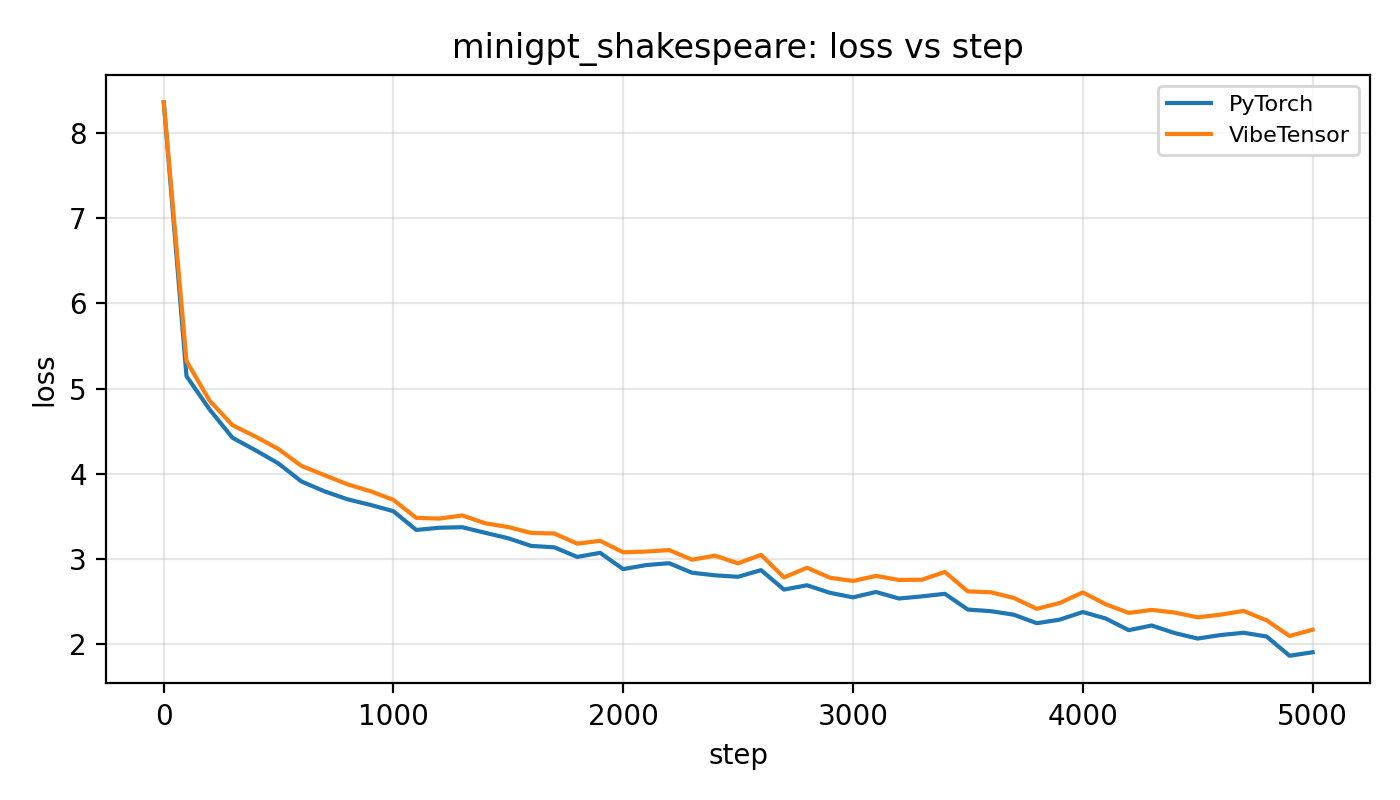}\\
    \small (c) miniGPT: loss
  \end{minipage}

  \vspace{0.5em}

  \begin{minipage}{0.32\textwidth}
    \centering
    \includegraphics[width=\linewidth]{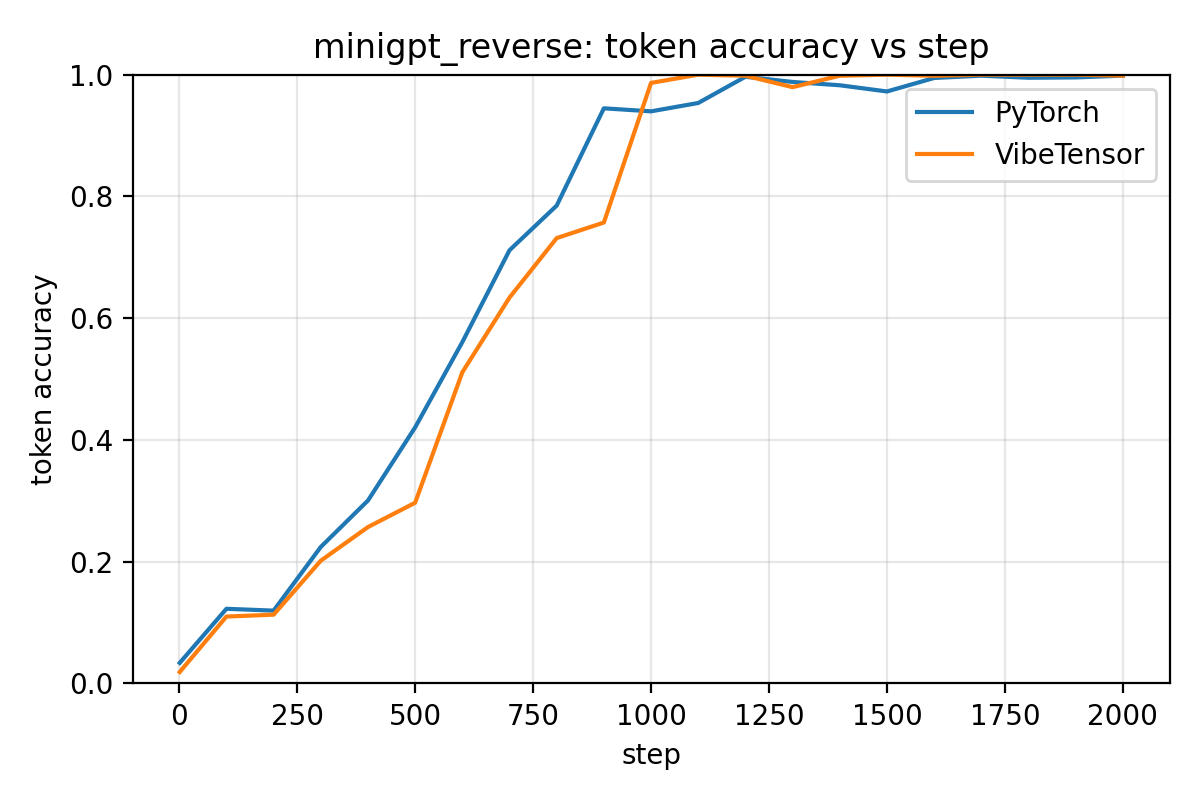}\\
    \small (d) Reversal: token accuracy
  \end{minipage}
  \hfill
  \begin{minipage}{0.32\textwidth}
    \centering
    \includegraphics[width=\linewidth]{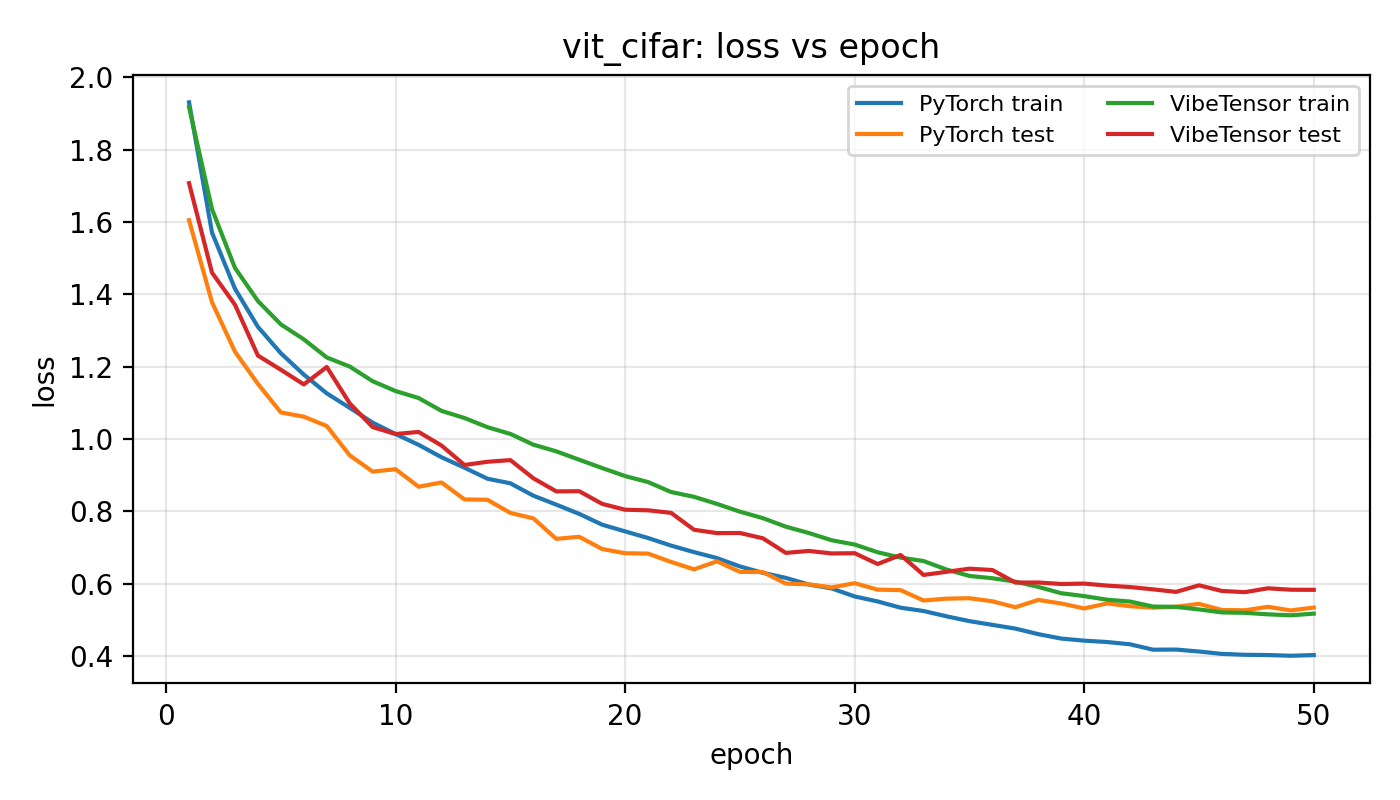}\\
    \small (e) ViT: loss
  \end{minipage}
  \hfill
  \begin{minipage}{0.32\textwidth}
    \centering
    \includegraphics[width=\linewidth]{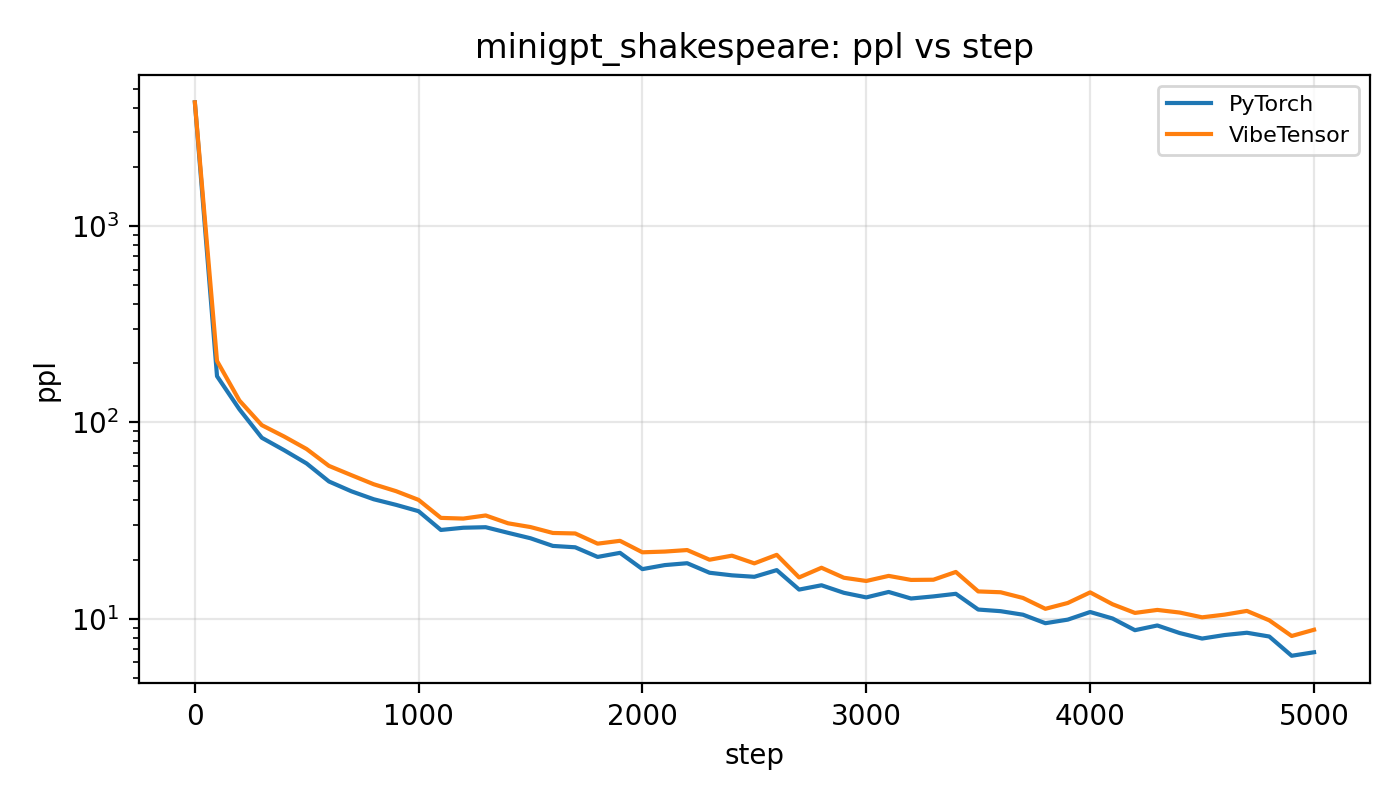}\\
    \small (f) miniGPT: perplexity
  \end{minipage}
  \caption{End-to-end training curves on Hopper H100 comparing \vt{} and PyTorch across three workloads. In each case, \vt{} matches the qualitative convergence behavior of the PyTorch baseline (loss decreases; accuracy/perplexity improve), indicating that core tensor semantics, autograd, and optimizer updates compose correctly at training-loop scale.}
  \label{fig:training_curves}
\end{figure*}

Beyond unit tests and isolated microbenchmarks, we validated that \vt{} composes into full training loops by running three small workloads: (i) a synthetic sequence-reversal task with a 2-layer decoder-only Transformer, (ii) a CIFAR-10 Vision Transformer based on Compact Vision Transformers~\cite{cvt}, and (iii) a miniGPT-style language model~\cite{mingpt} on Shakespeare.
We ran these workloads on both NVIDIA H100 (Hopper) and Blackwell GPUs to sanity-check cross-architecture behavior.
Figure~\ref{fig:training_curves} plots representative training curves from the Hopper H100 runs; Table~\ref{tab:training} summarizes mean timing across both architectures.
Across these workloads, \vt{} is currently slower than PyTorch (1.7--6.2$\times$), consistent with its prototype status and with known serialization points discussed in Section~\ref{sec:limitations}.
These results are intended as end-to-end functional checks rather than performance claims.

\begin{table*}[t]
  \caption{End-to-end training workloads and mean timing (recorded in release artifacts). Iteration times exclude the first logged step where noted; epoch times are averaged over the compared span.}
  \label{tab:training}
  \centering
  \small
  \setlength{\tabcolsep}{4pt}
  \begin{tabular}{p{0.33\textwidth} p{0.20\textwidth} r r r}
    \toprule
    Workload & Span (GPU) & PyTorch & \vt{} & Slowdown \\
    \midrule
    Sequence reversal (2-layer Transformer) & 2000 steps (Hopper H100) & 3.958 ms/iter & 12.02 ms/iter & 3.04$\times$ \\
    Sequence reversal (2-layer Transformer) & 2000 steps (Blackwell) & 7.25 ms/iter & 12.48 ms/iter & 1.72$\times$ \\
    CIFAR-10 (ViT, depth 6) & 50 epochs (Hopper H100) & 6.544 s/epoch & 37.67 s/epoch & 5.76$\times$ \\
    CIFAR-10 (ViT, depth 6) & 14 epochs (Blackwell) & 5.585 s/epoch & 34.36 s/epoch & 6.15$\times$ \\
    Shakespeare (miniGPT, 6 layers) & 5000 steps (Hopper H100) & 13.94 ms/iter & 80.62 ms/iter & 5.79$\times$ \\
    Shakespeare (miniGPT, 6 layers) & 5000 steps (Blackwell) & 18.63 ms/iter & 74.81 ms/iter & 4.01$\times$ \\
    \bottomrule
  \end{tabular}
\end{table*}

\subsection{Multi-GPU scaling (Fabric + ring allreduce backend)}
\begin{figure}[t]
  \centering
  \begin{minipage}{0.49\linewidth}
    \centering
    \includegraphics[width=\linewidth]{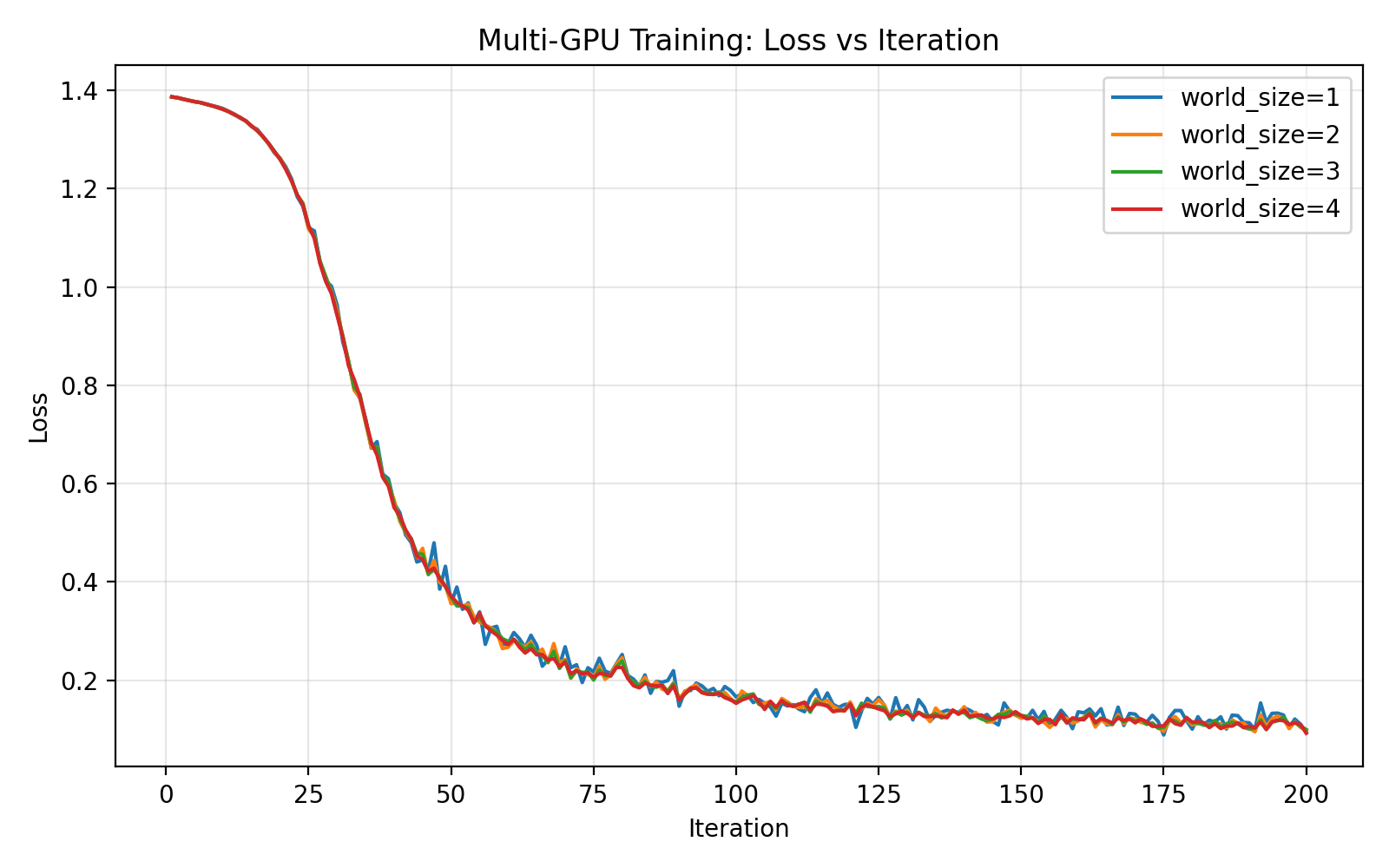}\\
    \small (a) Loss
  \end{minipage}
  \hfill
  \begin{minipage}{0.49\linewidth}
    \centering
    \includegraphics[width=\linewidth]{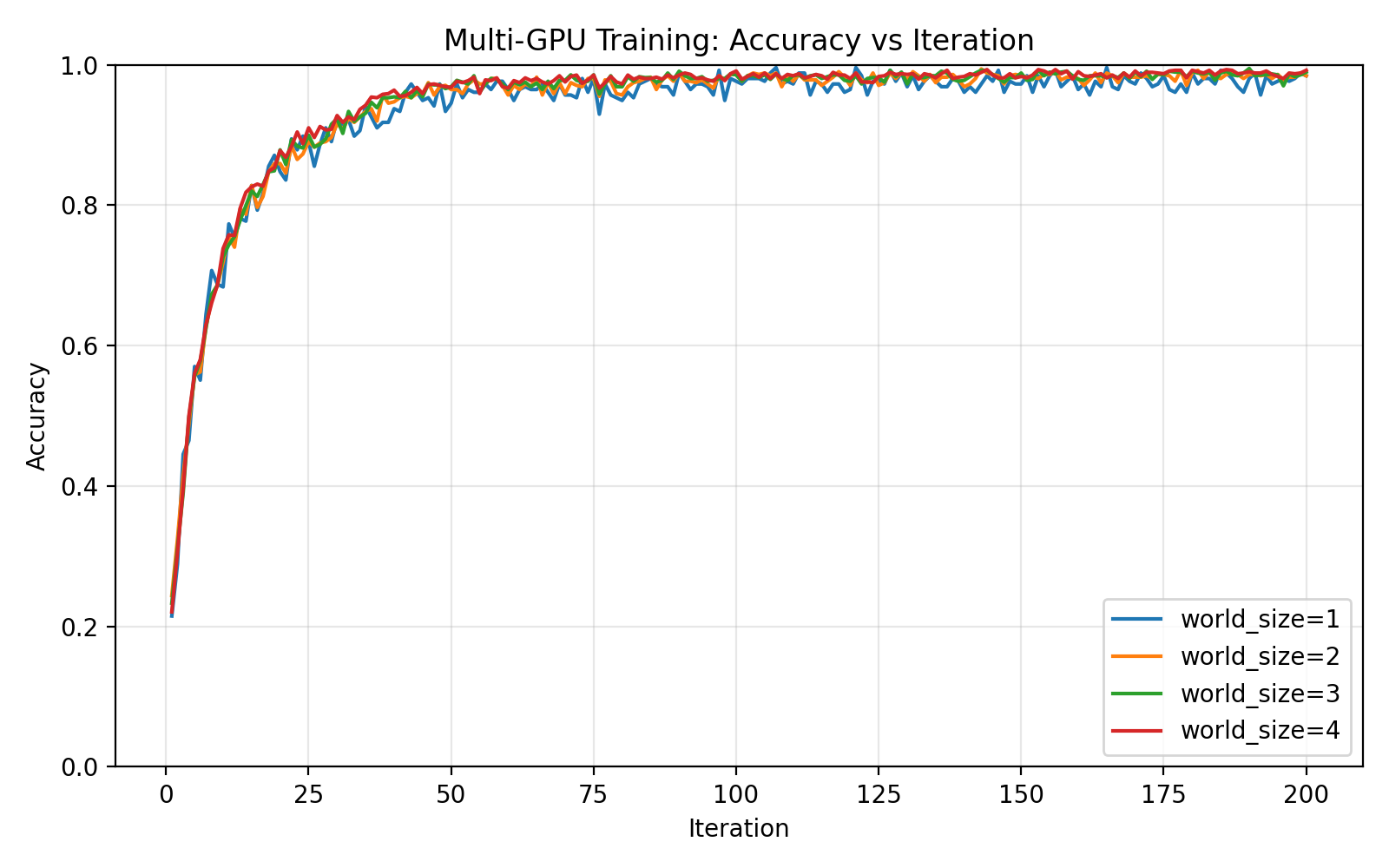}\\
    \small (b) Accuracy
  \end{minipage}
  \caption{Multi-GPU training curves for the Blackwell scaling benchmark (world size 1--4).}
  \label{fig:multigpu_curves}
\end{figure}

We also report a small multi-GPU benchmark on Blackwell GPUs using the experimental Fabric subsystem (Section~\ref{sec:fabric}) with a ring-allreduce backend (\code{fabric\_ring}).
Figure~\ref{fig:multigpu_curves} shows representative training curves, and Table~\ref{tab:multigpu} shows throughput scaling to four GPUs under weak scaling (fixed per-GPU batch).
Because this pathway depends on a Blackwell-only CUTLASS plugin, it was not run on the Hopper system.
Fabric is not a full distributed training runtime; these results primarily indicate that cross-device synchronization and observability pathways execute end to end.

\begin{table}[t]
  \caption{Multi-GPU training benchmark using \vt{} Fabric on Blackwell GPUs (weak scaling).}
  \label{tab:multigpu}
  \centering
  \small
  \begin{tabular}{rrrr}
    \toprule
    world\_size & per-GPU batch & avg iter (ms) & throughput (samples/s) \\
    \midrule
    1 & 65536 & 29.88 & $2.19\times 10^6$ \\
    2 & 65536 & 43.65 & $3.00\times 10^6$ \\
    3 & 65536 & 60.36 & $3.26\times 10^6$ \\
    4 & 65536 & 70.98 & $3.69\times 10^6$ \\
    \bottomrule
  \end{tabular}
\end{table}

\section{Limitations and lessons}
\label{sec:limitations}
\vt{} is a research prototype and carries substantial limitations.

\begin{figure}[t]
  \centering
  \includegraphics[width=\linewidth]{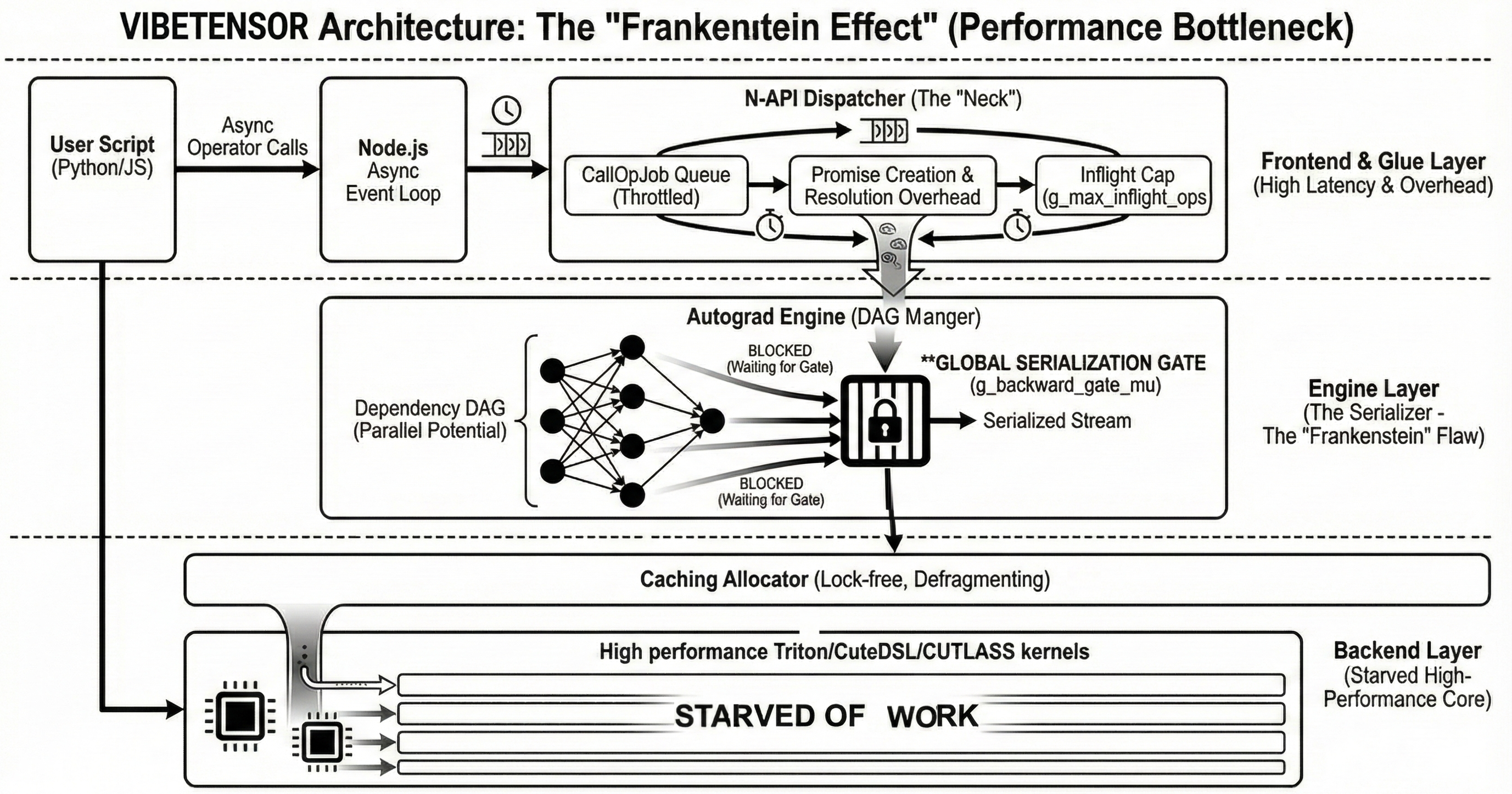}
  \caption{An example ``Frankenstein'' effect: a high-overhead frontend/engine layer can serialize execution and starve a high-performance backend. This reflects a limitation of correctness-first generation when global performance objectives are not encoded early.}
  \label{fig:frankenstein}
\end{figure}

\paragraph{The ``Frankenstein'' composition effect.}
A recurring failure mode in generated systems is that individually reasonable components can compose into a globally suboptimal design.
In \vt{}, a correctness-first autograd and dispatch design can introduce serialization points that starve otherwise efficient backend kernels.
One example is a non-reentrant global backward gate used to reject concurrent multithreaded backward runs (implemented as a process-wide try-locked mutex).
It simplifies safety properties (e.g., avoids deadlock patterns from nested backward) but can serialize independent backward work.
Figure~\ref{fig:frankenstein} walks through this bottleneck: first a global gate in the engine funnels concurrent backward calls; then host-side serialization and synchronization propagate to the device; finally, otherwise efficient backend kernels observe reduced utilization due to starvation.

\paragraph{Incomplete API surface and performance.}
\vt{} intentionally does not aim for full PyTorch compatibility.
Many operators, datatypes, and distributed features are missing or incomplete, and performance has not been tuned to match production frameworks.

\paragraph{Validation gaps unique to generated code.}
Agent-generated code can pass local unit tests while failing under repeated composition (e.g., multi-step training loops) due to stateful interactions, uninitialized buffers, or accidental global synchronization.
This motivates regression tests that exercise repeated execution, cross-stream behavior, and long-running workloads.

\paragraph{Maintenance, safety, and security.}
Machine-generated code can include inconsistent conventions, redundant abstractions, and subtle correctness or security issues.
We therefore caution against production use and position \vt{} primarily as a research and educational artifact.

\section{Conclusion}
\vt{} demonstrates that AI-assisted workflows can generate a non-trivial, GPU-enabled deep learning system software stack spanning Python and Node.js frontends and a C++20/CUDA core, validated by builds and tests.
The open-source release is intended to support reproducible study of system-scale software generation: what can be generated quickly, what kinds of bugs and bottlenecks arise, and what testing and architectural scaffolding most effectively constrain the search space.

From an ecosystem perspective, releasing such artifacts can benefit the broader GPU software community by providing concrete, hackable implementations of runtime and kernel components and by enabling rigorous discussion of AI-assisted engineering methods grounded in source code and tests.

\section*{Acknowledgments}
We thank the maintainers of foundational open-source projects referenced or used by \vt{}, including PyTorch, DLPack, nanobind, Triton, and CUTLASS.
We also thank internal reviewers who provided feedback on the generated system and draft.

{\small
\bibliographystyle{unsrtnat}
\bibliography{references}
}

\end{document}